# Air ions induced aerosol sensing by eye-safe LIDAR

[1]S.M. Pershin, [1]V.A. Zavozin, [1]M.Ya. Grishin., [1]V.N. Lednev, [1,2]G.A. Boldin, [3]L.B. Bezrukov, [3]A.K. Mezhokh, [2,3]V.V. Sinev

[1] A.M. Prokhorov General Physics Institute of Russian Academy of Sciences, Moscow, Russia

[2] National Research Nuclear University MEPhI, Moscow, Russia

[3] Institute for Nuclear Research, Moscow, Russia

*Low concentrations aerosols quantification is rather challenging for LIDAR instruments due to eye-safety restrictions so high energy pulses cannot be utilized to improve the sensitivity. Highly sensitive but eye-save LIDAR has been developed for the quantification of the water droplet aerosol which was induced by air ions. Few days sensing of aerosols in closed tunnel revealed a strong correlation between air optical transparency (LIDAR measurements) and concentrations of positive/negative ions (ion counter Sapphir 3-M). The correlation coefficient was observed to be almost unity for the air transparency signal and air ions unipolarity coefficient. High sensitivity of the water droplet aerosol quantification makes the developed eye-safe LIDAR a perspective instrument for space resolved measurements of the air ions distribution. Space and time resolved measurements of air ions exhalation can be a new instrument for tectonic activity study including new earthquake forecasting indicators search.*

## Introduction

Earthquake forecasting is an ongoing debate within the last century in the scientific community. Earthquake early warning can be made a few seconds in advance based on P-wave detection [1] but a few hours are needed to alarm the citizens thus preventing loss of lives. Numerous reports of both seismic and non-seismic signals measured before earthquakes are appeared every few years but nature of these signals required proofs. Still, developing new techniques for early detection of tectonic activity is of high importance though reliable indicators of eruptions or earthquakes are still absent [2–4]. Several signals from few minutes to few hours have been published in the literature during last 50 years : electromagnetic signal in kHz to MHz frequencies; P and S wave velocities comparison [5]; radon gas exhalations [6]; atmospheric and ionosphere anomalies [7–10].

Careful review on radon gas exhalations anomalies publications revealed an existence of radon emission abnormalities before the earthquakes but these features look strikingly similar to non-tectonic ones [8]. Still, searching new pre-earthquake indicators is of high interest and new attempts are published. For example, high concentrations of air ions and 3-fold aerosol concentration growth have been detected 2 hours before the earthquake [11]. The modern explanation of the increased air ions concentrations is still under the debates [12–15]. For example, Freund et al. [16] introduced the electronic charge carrier theory to explain the observed increase of the ions concentration. Alternatively, Pulinets et al. suggested that increased radon exhalation triggered the increased concentrations of the ions [9,17]. Additional gases emission monitoring (helium, hydrogen, $H_2O$, $SO_2$ and $CO_2$) have been found generally to be increase before the tectonic activity growth thus suggesting to be a new source of information on tectonic processes [18–22]. For example, Stromboli volcano (Italy) emission of $CO_2$ and $SO_2$ gases and its ratio significantly changes few hours before the tectonic activity [23]. Recently, Ghosh et al. [13] demonstrated that 3-7 days before the earthquake an abnormal aerosols formed near epicenters of Kumamoto and Fukushima earthquakes.

The magmatic gasses emitted trough the small holes and cracks network in the rocks and it is difficult to detect visually. If sensors will be place far away from the degassing spots then reliability of magmatic gases monitoring can be distorted. Moreover, degassing spots can change it locations during tectonic activities as well as impacted on the degassing fluxes. One way to solve this problem is to construct a huge network of air ion sensors to get the meaningful results as have been shown by Warden et al. [24]. Alternative was to track all the degassing gases is to use a remote sensing techniques like laser remote sensing instruments LIDARs (Light Detection And Ranging). For example, LIDAR technology was successfully utilized for remote sensing of different gases at volcano vent [19,23].

Recently, we have demonstrated that aerosols dynamics in the closed tunnel near the volcano have strong correlation with the Earth crust deformation (measured by laser strain meter). A long term monitoring within the seasons have demonstrated a good correlation between the aerosol concentration fluctuations and Earth crust deformations [25–27]. However, a few significant fluctuations of aerosol signal were detected while no Earth crust deformation, air pressure and temperature have been measured [11].

The abrupt aerosol growth occurred 2 hours and 10 minutes before the 6.1 M earthquake at Luzon Island (Philippines). Supposing that radon exhalation through the holes and cracks in the tunnel walls will result in air ions concentration growth in the tunnel. Increased air ions concentration will trigger the water droplet aerosol formation. Here, we decided to systematically study the impact of air ions on aerosol formation inside the isolated underground tunnel. The first goal is to quantitatively estimate the how air ions fluctuation can impact on aerosol concentrations in order to explain the unknown aerosol concentrations spikes during long-term monitoring in close tunnel near Elbrus. The second purpose of the study is to estimate sensitivity of eye-safety LIDAR sensing for detecting small variations of the aerosols induced by air ions.

## Experimental part

In order to check the above mentioned correlation between water aerosol density and air ion concentration we have choose the location of the experiment to fulfill the following: spot should be closed from the sunlight to prevent UV photons to generate air ions; the location should be closed but air ventilation capability is required to blow off the air ions; it is highly desirable that slow generation of air ions takes place in the location LIDAR and ion meter so sampling rate was greater the air ions variations. Keeping these requirements in mind we have chosen the underground laboratory room located 12 meters below ground level at the 3 level of the Lomonosov MSU Physical department building basement (Latitude 55.69 N, Longitude 37.54 E). The radon gas exhalation is generally slow in the basement so air ions generation will be also slow. The room can be ventilated with the fresh air upon request or time schedule. The scheme of the experiment location with the two instruments for sensing aerosol and air ions concentrations are shown in Fig. 1.

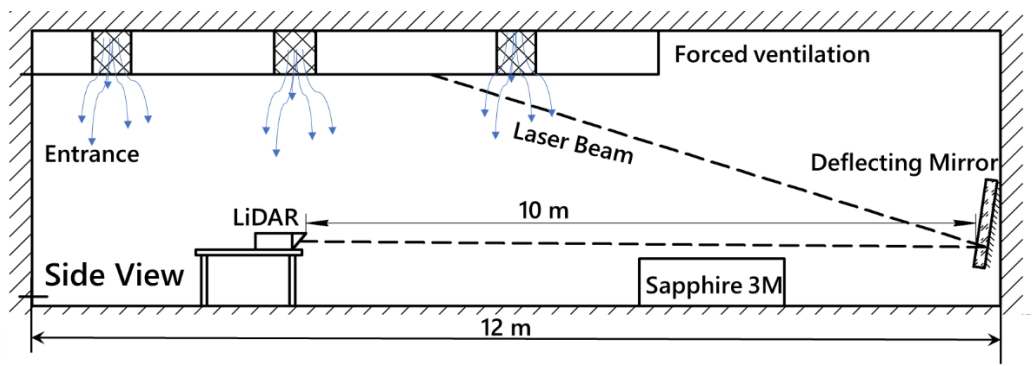

**Fig. 1** The schematic view of the experiment on simultaneous aerosol (by eye-safe LiDAR) and air ions sensing (by "Sapphire 3M" air ion counter) inside the closed tunnel.

Air ions concentrations were quantified by the modified version of the "Sapphire 3 M" instrument. The improved version of the air ion counter provided better reproducibility and accuracy of the air ions concentration measurements under high humidity conditions [28]. The instrument was remotely controlled via computer so the air ions data were acquired and stored automatically.

Aerosol was quantified by the eye-safe LiDAR developed in our laboratory for the high sensitivity measurements of the aerosols. The detailed description can be found in [27]. Shortly, the LiDAR is based on low energy pulsed diode laser and highly sensitive single photon avalanche photodiode thus fulfilling the eye-safe conditions. The physical principle of aerosol sensing is based on elastic scattering of the laser pulse by aerosol particles. The LiDAR was installed 2 m away of from the room entrance and the laser beam was directed to the opposite wall. We added an additional optical mirror to reflect the laser beam to the room ceiling since LiDAR provided space resolved measurements so the aerosol dynamics in the bottom and the top levels can be studied. The LiDAR instrument was controlled by computer and acquired data automatically according to the experiment program schedule. Air ion meter was installed near the laser beam in the middle of the room 0.5 m above the floor level. Due to the fundamentals of the eye-safety LiDAR operation [27] and to get good signal-to-noise ratio the instrument provided 20 000 laser pulses every 10 minutes (laser rate is 4 000 Hz so a single measurement takes 5 seconds). The "Sapphire 3 M" instrument measured air ions concentrations every 10 minutes. The LiDAR and air ion meter was synchronized so the variation of the aerosol and air ion concentration can be compared in time. Air temperature, relative humidity and pressure inside room were measured by sensors: relative humidity was measured by Honeywell HIH-4000-003 with accuracy ±3.5%RH; air temperature - Analog Devices AD22100K with accuracy ±0.5 °C; air pressure - Freescale Semiconductor MPXA4100A6U/T1 with accuracy ±11.25 mmHg.

## Results and Discussion

The LiDAR provided the photon counting histogram during single measurement and several signals can be as shown in Fig. 2. The first peak in the histogram at 0 to 8 m distance was attributed to the aerosol backscattering (grey color) so the photocounts were summed and defined as an aerosol backscattering signal. The mirror backscattering has not been used in the study since it is originated from the mirror surface reflection mainly due to the deposited dust. The aerosol backscattering between mirror and back wall was very small for reliable measurements so we cannot accept it as a signal. The third peak is due to the back wall located near the LiDAR (see Fig. 1) so it was defined as the "trace transparency" signal on the round trip pass since this signal is proportional to the air transparency (the clear the air in the tunnel the greater the wall backscattering signal will be). During each measurement we also defined the noise of the measurement as the photocounts integral at 30 to 40 m distance (not shown in Fig. 2) in order to check the LiDAR operation. It should be noted that both laser and detector provided the same noise level during all the period of the measurements.

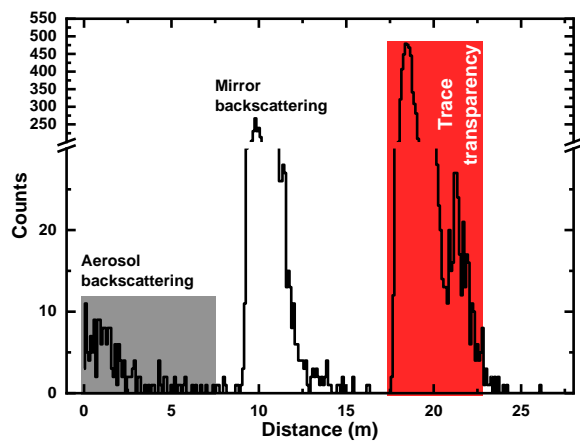

**Fig. 2** Photon histogram acquired by LiDAR along the sensing path. Aerosol scattering signal defined by the photon counts integrated within the first 8 meters distance (grey color). The mirror and trace transparency signals were located at 10 and 20 m distance respectively from the LiDAR so the corresponding backscattering signals were defined accordingly (white and red color).

According to the experiment goal the air ions concentrations should be varied so we turned on the ventilation in the room every morning in order to fill the room with the fresh air. After midday the ventilation was switched off so the radon and additional positive ions (hydroxonium) were slowly concentrated in the room. The air humidity began to increase also. A data series during six days measurements (5-11 December 2022) of the aerosols (by LiDAR) and air ions (by Sapphire-3M) are presented in Fig. 3. The periodic changes of the signals are due to the air ventilation. The aerosol backscattering signal (black color in Fig. 2) was low level (~3200 photocounts on 20000 laser pulse) but small variations can be traced in Fig.3. The aerosol backscattering signal didn't correlate with the air ions concentrations. Alternatively, the trace transparency signal showed negative correlation with the air ion concentrations as well as with the aerosol backscattering signal. Previously, we have demonstrated that air transparency measurements by LiDAR is a better choice for low concentration aerosol quantification rather than direct measurements of the aerosols backscattering [11]. It can be clearly observed in Fig. 3 (c) that positive ions concentration was systematically greater the concentration of negative ions.

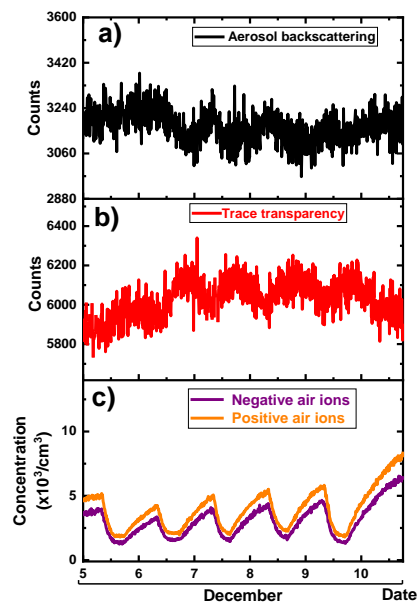

Fig. 3 Time sets series of the LiDAR and air ions counter measurements during 5 days period: (a) – aerosol backscattering signal (black color); (b) – trace transparency signal (red color); (c) – positive (orange) and negative (violet) air ion concentrations.

In order to make the better view of the results, we have filtered the trace transparency signal by Savitsky-Golay smothering (Fig. 4) [29]. Such anti-aliasing polynomial smothering is frequency utilized in laser physics due better tendency to secure the high frequency signal components compared to the conventional averaging by non-recursive filters [30]. Data was filtered with the 35 data point window and 5 order of the fitting polynomial function. A better negative correlation between wall backscattering signal and air ions concentrations can be observed after smoothing. It should be noted that the trace transparency signal delayed compared to the beginning of increase in the ionic concentrations.

It should be noted that positive and negative air ions ratio in Fig. 4 (c) was not the constant during the fresh air ventilation. Typically, a unipolarity coefficient is used to define positive to negative ions concentration ratio for two reasons [24]: first, parameter had tendency to change abruptly before the tectonic events; second, comparing unipolarity coefficient with the positive/negative ions concentrations the malfunction of the sensor can be detected. The unipolarity coefficient data were also filtered by Savitsky-Golay smoothening. The ventilation was less effective for positive air ions compared to the negative ions so the unipolarity coefficient was growing while ventilation was switched on. When ventilation was stopped the unipolarity decreased. A good correlation between LiDAR signal and unipolarity coefficient can be established according to the Fig. 4 (a) and (c). The major source of air ions is the radon which had high solubility to the water. Air humidity was decreased as the fresh air was ventilated inside the room so relative humidity was also quantified and is shown in Fig. 4 (d). The air temperature was almost constant during the measurements and also did the air pressure (see in supplementary materials Fig. S.1). Comparing LiDAR signal, unipolarity coefficient and air humidity data series a good correlation can be established. This correlation can be explained by water aerosol generation induced by air ions. Interestingly, but during observation period the relative humidity had tendency to decrease while trace transparency signal and unipolarity coefficient were increasing (Fig.4).

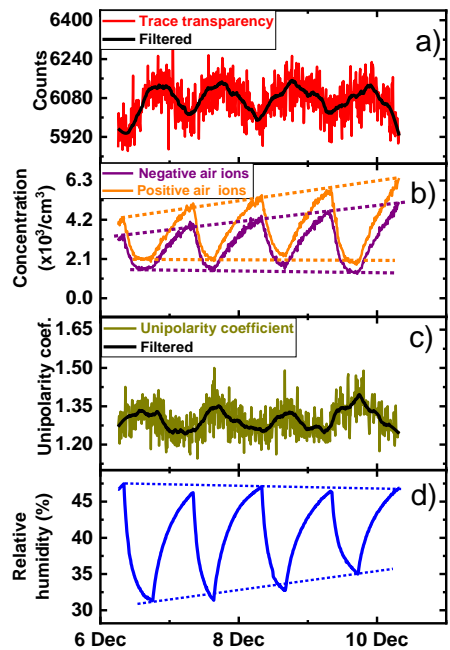

Fig. 4 The aerosol and air ion sensing signals series during 5 days period: (a) – raw LiDAR's trace transparency signal (red) and after filtering (black); (b) – positive (orange) and negative (violet) air ions concentrations; (c) – air ions unipolarity coefficient for raw data (brown) and after filtering (black); (d) – relative humidity. The air temperature and pressure values were constant during the measurements and the corresponding data was presented in supplementary materials Fig. S.1.

For better comparison we have constructed correlation plots for positive and negative ions as well as its ratio as the function of the LiDAR signal (Fig. 5). Spirman's correlation coefficient calculated for all the data points was rather high (greater 0.7) in all the cases but better correlation was established for unipolarity coefficient. It should be noted that some data points had almost linear correlation between air ions concentration and the LiDAR signal (see left part in the correlation plot data points). The data points for this part were extracted to find out the measurements when the correlation is the strongest and result is presented in Fig. 6.

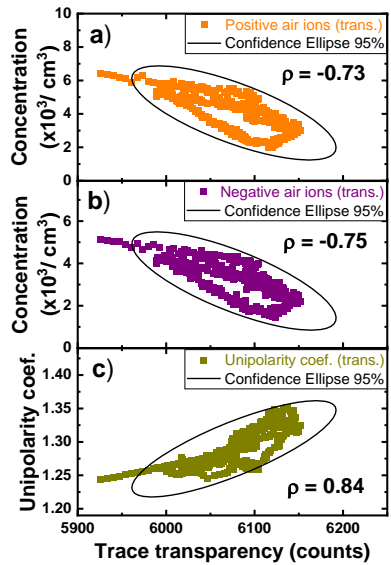

Fig. 5 The correlation plots for the wall backscattering signal (inversely proportional to the air transparency): (a) – positive ion concentration vs trace transparency signal (orange color); (b) – negative ion concentration vs trace transparency signal (violet color); (c) – unipolarity coefficient vs trace transparency signal (brown color). Spirman's correlation coefficient (ρ) was plotted inlet. The black solid ellipse depicts the correlation with the 0.95 probability level.

According to the Fig. 6 the Spearman's rank correlation coefficient between air ion concentrations and the trace transparency was almost unity (ρ=0.9997). This data points corresponds to the moment when ventilation was switched off during the night. Consequently, during period of natural radon concentration growth in the underground tunnel the concentration of ions is growing resulting in water aerosol formation (decreasing trace transparency, the beginning scale in Fig.5 and Fig.6). The ventilation partially hided the radon exhalation so the LiDAR signal had less correlation with the air ions concentrations.

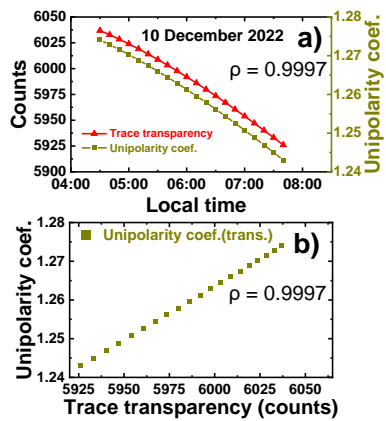

Fig. 6 (a) - The detailed plots for the LiDAR signal (proportional to the air transparency, red color) and unipolarity coefficient (brown color) data series. (b) - The correlation plot for unipolarity coefficient *vs* trace transparency signal (LiDAR signal).

## Conclusions

The detailed study of aerosol and air ions concentrations in the underground lab revealed a strong negative correlation between air transparency and air ions concentrations. The air transparency was quantified by the highly sensitive but eye-safe aerosol LiDAR instrument while air ions concentrations were monitored by precise counter Sapphir 3-M. Strong correlation (Spirman's coefficient $\rho=0.9997$) can be explained by the low concentration water droplet aerosol generation induced by the air ions. Under ventilation the correlation between LiDAR signal and air ions measurements became more poor but it can be still established ($\rho=0.73$). High sensitivity of the water droplet aerosol quantification makes the developed eye-safe LiDAR a perspective instrument for large scale and space resolved measurements of the air ions distribution near the Earth surface or inside the Baksan Neutrino Observatory tunnel [29] so the radon exhalation spots can be detected in real time. It is believed that the developing network of the LiDAR instruments will provide more reliable results on radon exhalation so the tectonic activity can be traced precisely.

## Declaration of competing interest

The authors declare that they have no known competing financial interests or personal relationships that could have appeared to influence the work reported in this paper.

## Acknowledgements

The authors gratefully acknowledge the Russian Science Foundation for financial support of the study (Project No. 19-19-00712). Authors thanks Yu.V Stenkin and L.A. Kuzmichev for help with the experiments in the underground lab at SINP MSU.

## Conflict of interest

The authors declare no conflicts of interest.

# Supplementary materials

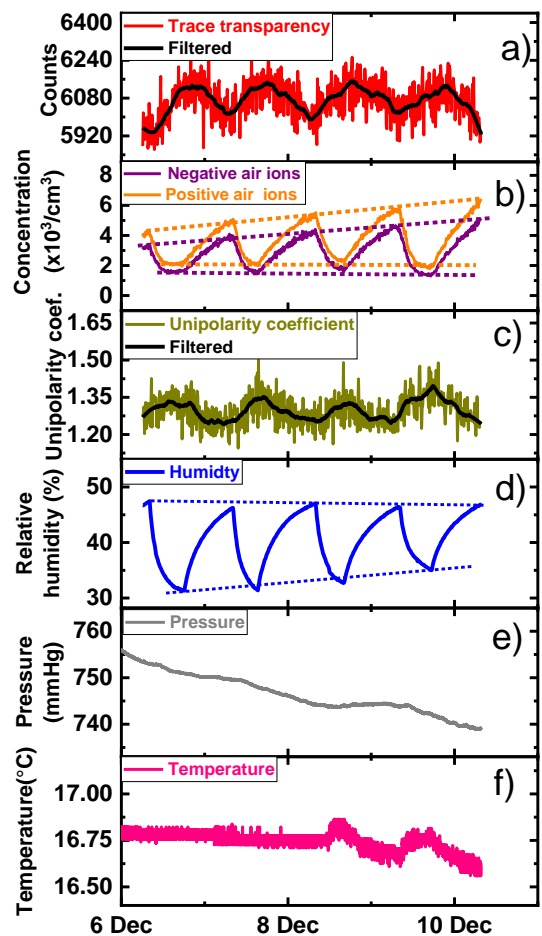

**Fig. S.1** The aerosol and air ion sensing signals series during 5 days period: (a) – raw LiDAR's trace transparency signal (red) and after filtering (black); (b) – positive (orange) and negative (violet) air ions concentrations; (c) – air ions unipolarity coefficient for raw data (brown) and after filtering (black); (d) – relative humidity (Honeywell sensor HIH-4000-003, accuracy ±3.5%RH); (e) - air temperature (Analog Devices sensor AD22100K, accuracy ±0.5 °C); (f) – air pressure (Freescale Semiconductor sensor MPXA4100A6U/T1, accuracy ±11.25 mmHg).